\documentclass[11pt,fleqn]{elsart}

\usepackage{amsmath,amssymb}
\usepackage{exscale}
\usepackage{longtable,multirow}

\allowdisplaybreaks[1]

\newfont{\hermes}{cmtt10 at 10pt}

\newcommand{\fs}{\; .}
\def\co{\; ,}

\renewcommand{\theequation}{\arabic{equation}}

\newcommand{\mk}{\bar{M}_K}

\renewcommand{\L}{\mathcal{L}}
\newcommand{\osn}[1]{\oldstylenums{#1}}

\newcommand{\qcd}{\sc qcd\rm}
\newcommand{\eom}{\sc eom\rm}
\newcommand{\chpt}[1]{$\chi\text{\sc pt\rm}_{#1}$}
\newcommand{\lecs}{\sc lec\rm s}
\newcommand{\lec}{\sc lec\rm }
\newcommand{\SU}[1]{\mathrm{SU}(#1)}
\renewcommand{\d}{\partial}

\newcommand{\Lr}[1]{L^\mathrm{r}_{#1}}
\newcommand{\lr}[1]{l^\mathrm{r}_{#1}}
\newcommand{\Hr}[1]{H^\mathrm{r}_{#1}}
\newcommand{\hr}[1]{h^\mathrm{r}_{#1}}
\newcommand{\clr}[1]{c^{\mathrm{r}}_{#1}}
\newcommand{\Clr}[1]{C^{\mathrm{r}}_{#1}}

\begin{document}

\renewcommand{\theequation}{\arabic{equation}}
\begin{frontmatter}

\title{\Large\bf Integrating out strange quarks in ChPT: terms at order $p^6$ }

\author[Bern]{J.~Gasser},
\author[Bern]{Ch.~Haefeli},
\author[Dubna]{M.A.~Ivanov}
\author[Bern]{and M.~Schmid}
\address[Bern]{Center for Research and Education in Fundamental Physics,
\\Institute for Theoretical Physics, University of Bern,
Sidlerstr. 5,
\\CH--3012 Bern, Switzerland}
\address[Dubna]{Laboratory of Theoretical Physics,
Joint Institute for Nuclear Research, \\
141980 Dubna (Moscow region), Russia}

\begin{abstract}
Chiral perturbation theory in the two--flavour sector 
allows one to analyse Green functions in \qcd{} in a limit where 
the strange quark mass is considered to be large in comparison to the 
external momenta and to the light quark masses $m_u$ and $m_d$.
In this framework, the low--energy constants of $\SU{2}_R\times \SU{2}_L$ depend 
on the value of the heavy quark masses. In a recent article, we have 
worked out, 
for the coupling constants $l_i$ 
which occur at order $p^4$ in the chiral expansion,
 the dependence on the strange quark mass at two--loop accuracy.
Here, we provide analogous relations for some of the  couplings $c_i$ which 
are relevant at order  $p^6$. To keep the calculations 
 somewhat reasonable in size, we consider only 
those \osn{28} couplings which enter the Green functions
 built from vector- and axial vector 
quark currents in the chiral limit $m_u=m_d=0,m_s\neq 0$. This provides the matching 
for \osn{27}
linear combinations of the \osn{28} couplings.
\end{abstract}


\end{frontmatter}


\section{}
\vspace{-2\parskip}
At low energies and  small quark masses, the Green functions of quark currents  can be analysed in the 
framework of chiral perturbation theory (\chpt{})\cite{weinberg,glann,glnpb}. The method allows
 one to work out the  momentum and quark mass dependence of the quantities of interest in a 
systematic and coherent manner.
It is customary to perform the quark mass expansion either around
$m_u=m_d=0$, with the strange quark mass held fixed at its physical value
(\chpt{2}), or to consider an expansion in all three quark masses around
$m_u=m_d=m_s=0$ (\chpt{3}).
The corresponding  effective Lagrangians contain  
low--energy constants (\lecs{}) that parametrise the degrees of freedom which are integrated out.
 The two expansions are not
independent:  
one can express the \lecs{} in the
two--flavour case through the ones in \chpt{3}, referred to as \emph{matching} in
the following. In Ref.~\cite{glnpb}, the pertinent relations for the couplings $l_i$ -- 
which occur at order $p^4$ in \chpt{2} -- were worked out at one--loop order.
 Recently, this matching has been performed  at two--loop  order~\cite{Gasser:2007sg}.

In this article, we investigate the analogous relations
for the \lecs{} $c_i$ which enter the effective Lagrangian of \chpt{2} at order $p^6$.  
The structure of the expansion is the following,
\begin{equation}
c_i=\frac{d_{i2}F^4}{\bar M_K^4}+\frac{d_{i1} F^2}{\bar M_K^2}+d_{i0}
+O(m_s)\,,\, i=1,\ldots,56\fs
\end{equation}
Here, $F$($\bar M_K$) denotes the pion decay constant (the kaon mass) 
 in the chiral limit\footnote{Throughout this article, we denote by {\it chiral limit} the case
 where $m_u=m_d=0, m_s\neq 0$.}. 
 The constants $d_{im}$ are the 
coefficients we are after: the $d_{i2}, d_{i1},d_{i0}$ require a tree, one--loop and two--loop 
calculation, in order. Furthermore, the $d_{i0}$ are linear in the $p^6$ couplings $C_i$ 
{}from \chpt{3}. This shows that, in order to have the relation between the $c_i$ and $C_i$ at leading order correct, a two--loop evaluation of the local terms in the 
effective action of \chpt{3} 
 at order $p^6$ is needed. For  the corresponding  relations between $l_i$ and $L_i$ at 
leading order, the expansion of the one--loop action of \chpt{3} at order $p^4$ suffices.

It turns out that the required calculations are very complex. We circumvent the 
problem at the cost of loosing some information: we confine ourselves to the investigation of 
those \lecs{} that occur in the Green functions of vector- and axial vector currents 
in the chiral limit. 
This allows one to remove the external scalar and pseudoscalar sources in the effective Lagrangian of
 the three flavour framework nearly altogether: it suffices to set $s$=diag(0,0,$m_s$), 
and $p=0$. This simplifies the calculations 
considerably. On the other hand, as we will see, we can provide the matching for only \osn{27} 
linear combinations of the \osn{28} \lecs{} that occur in the Green functions mentioned.

We comment on related works  available in the literature, aside from the 
ones already mentioned. 
i) The strange quark
mass expansion of the \chpt{2} \lec{} $B$ ($F^2B$) was provided at two--loop
accuracy in 
Ref.~\cite{Kaiser:2006uv} (\cite{Mouss:Sigma}). 
ii) Matching of the order $p^6$ 
\lecs{} in the parity--odd sector was performed recently in Ref.~\cite{kampf_moussallam}.
iii) Analogous work was done  in the baryon sector
in Ref.~\cite{meissner_frink}, and for electromagnetic interactions in
 Refs.~\cite{gasser_rusetsky,jallouli_sazdjian,NehmePhD,Haefeli:2007ey}. 
 iv) The authors of Refs.~\cite{GChPT,dcpipikk} 
investigate what happens
 if chiral symmetry breaking exhibits different patterns in \chpt{2} 
and \chpt{3}. The literature  on the subject 
may be traced from Ref.~\cite{dcpipikk}.
 In this scenario, a substantial strange quark mass dependence may 
show up, as a result of which \chpt{3} must be reordered
 and the effect of vacuum fluctuations of $\bar ss$ pairs 
 summed up.   
 Whether the relations provided below
favour such a situation is not investigated here -- 
 the present work just 
provides the {\it algebraic} dependences of the \chpt{2} \lecs{} on the
 strange quark mass, at two--loop order.

The article is structured as follows.
In section~\ref{sec:ex}, we illustrate the matching considered here with the
vector--vector correlator, where a two--loop calculation in \chpt{3} is available \cite{Amoros:1999dp}, 
and where the two--loop result of \chpt{2} can be  read off easily, as a  result of which the 
matching becomes almost trivial. In section~\ref{sec:method}, we turn to the general
case and outline the method used. 
In section~\ref{sec:result}, we first present a linear combination 
between the local polynomials at order $p^6$ 
which holds in the restricted framework.
Afterwards, we display the matching relations. 
The final section~\ref{sec:summary}  contains concluding remarks.

\section{}
\label{sec:ex}
\vspace{-2\parskip}
In principle,  a matching of the \lecs{} can be achieved by comparing 
pertinent matrix elements, calculated in both theories up to two--loop order,
 in the chiral limit: One then simply needs to expand the amplitudes 
from \chpt{3} in
small momenta up to the relevant order and establish the
relations between the $\SU{2}$-- and $\SU{3}$--\lecs{} by equating
the results of the two calculations. To illustrate the procedure, consider the matching
in the case of the vector--vector correlator $\Pi_{V\pi}$.
This quantity was
evaluated in the framework of \chpt{3} to two loops in 
Refs.~\cite{Golowich:1995kd,Amoros:1999dp}.
In the chiral limit, the corresponding expression
 in the framework of \chpt{2}  may be easily obtained from the
three flavour one by dropping the kaon contributions
and replacing the $\SU{3}$--\lecs{} with the ones of $\SU{2}$,
 e.g.~$\Lr{9}\to -\frac{1}{2} \lr{6}$,
$\Clr{93}\to \clr{56}$,~etc. In momentum space,
\begin{gather}
\label{eq:VV_SU2}
\begin{split}
\Pi^{\SU{2}}_{V\pi}(t) = &8\hr{2}-4 B_{V\pi}(t)\\
&+\frac{t}{F^2}\big[8 B_{V\pi}(t)
\big(\lr{6} + B_{V\pi}(t)\big) - 4 \clr{56}\big]+O(F^{-4})\co
\end{split}\\
\begin{split}
\Pi^{\SU{3}}_{V\pi}(t)
=&-2( 2\Hr{1} +\Lr{10})-4B_{V\pi}(t) -2B_{VK}(t)\\
&+\frac{\mk^2}{F_0^2}
\big[ \tfrac{4}{N}\ell_K ( \Lr{9} +\Lr{10})
-32\Clr{62} \big]\\
&
+\frac{t}{F_0^2}\big[-8\big( 2B_{V\pi}(t) +  B_{VK}(t) \big) \Lr{9}\\
&\qquad\quad+2\big(2B_{V\pi}(t) +  B_{VK}(t) \big)^2 -4\Clr{93} \big]+O(F_0^{-4})\co
\end{split}
\end{gather}
where
\begin{equation}
\label{eq:defs}
\begin{split}
B_{VK}(t)&=\frac16 \nu_K-\frac{1}{120N}\frac{t}{\mk^2} +O(t^2)\co\\
\nu_K &=\frac{1}{2 N}(\ell_K+1)\co \quad \ell_K = \ln(\mk^2/\mu^2)\co\quad N=16\pi^2\fs
\end{split}
\end{equation}   
Here, $F_0$ denotes the pion decay constant $F$ at $m_s=0$.
The square of the kaon mass in the chiral limit reads \cite{glnpb}
\begin{equation}
\label{eq:Mk2}
\hspace{-0.5cm}\mk^2 = B_0 m_s\Big[1 
+ \frac{B_0 m_s}{F_0^2} \Big(\frac{4}{9 N} \ln\frac{4 B_0 m_s}{3\mu^2}
+16(2\Lr{6}-\Lr{4})+8(2\Lr{8}-\Lr{5})\Big)+O(m_s^2)\Big]\fs
\end{equation}
By requiring that $\Pi^{\SU{2}}_{V\pi}(t)=\Pi^{\SU{3}}_{V\pi}(t)$,
we find 
\begin{gather}
\label{eq:matching1}
\begin{split}
h^\mathrm{r}_2 = &-\tfrac12 H_1^\mathrm{r} -\tfrac14 \Lr{10} -\tfrac{1}{24}\nu_K\\
&+\frac{\mk^2}{F^2_0}
\big[ \tfrac{1}{2N}\ell_K (\Lr{9}+\Lr{10})
-4\Clr{62} \big]+O(m_s^2)\co
\end{split}\\
\label{eq:matching2}
\clr{56} = - \frac{1}{240N}\frac{F^2}{\mk^2}
            + \tfrac13 \nu_K \Lr{9} -\tfrac{1}{72}\nu_K^2 + \Clr{93}+O(m_s)\fs
 \end{gather}
The first line of Eq.~(\ref{eq:matching1}) was already derived in Ref.~\cite{glnpb},
whereas the terms proportional to $m_s$ in the second line have been
 calculated while working on \cite{Gasser:2007sg}. The result 
Eq.~(\ref{eq:matching2}) will be derived by a general method again below. Note
that the relation (\ref{eq:matching2}) involves a term proportional to
$1/m_s$, a situation similar to the case of $l_7$ \cite{glnpb}. There,
the singular term stems from a tree--level contribution, whereas for $c_{56}^r$,  it originates from the momentum expansion of the loop function
 $B_{VK}$ in Eq.~(\ref{eq:defs}).  Matching relations for the $c_i$ may involve 
terms proportional to $1/m_s^2$ as well. However,  these occur only in  monomials 
related to the sources $s$ and $p$.

We now outline a systematic method which allows one to obtain 
matching relations without the need to evaluate a large number of matrix elements.

\section{}
\label{sec:method}
\vspace{-2\parskip}
The idea is to restrict the physics of \chpt{3} to the one of \chpt{2}. To
this end, we impose the following restrictions, collectively referred to as
\emph{two--flavour limit:}
\begin{enumerate}
\item[i)] the external sources of \chpt{3} are restricted to the two--flavour
  subspace, with $m_s$ kept at its physical value;
\item[ii)] 
the matching is performed in the chiral limit;
\item[iii)] external momenta are restricted to values below the threshold of
  the massive fields, $|p^2| \ll M_K^2$.
\end{enumerate}
The matching relations can then be read off from equating the pertinent 
generating functionals in \chpt{3} and \chpt{2}.
[An analogous  method was established in Ref.~\cite{Nyffeler:1994ph} in the
context of the linear sigma model.] 

It is
straightforward to apply it at one--loop level to \chpt{} and to obtain
the relations presented in Ref.~\cite{glnpb} for the pertinent \lecs{}.
We have extended it to the two--loop level and established the
relations between the \chpt{2} \lecs{} -- which appear in the effective lagrangians $\L_2$ and $\L_4$ --
 and the corresponding \chpt{3} \lecs{} \cite{Thesis,Gasser:2007sg}. This technique
was also applied to determine the strange quark mass dependence 
of the electromagnetic two--flavour \lecs{} \cite{Haefeli:2007ey}.

In the present  work we do {not} deal with the full \chpt{}. Rather, we switch off
the sources $s$ and $p$ (while retaining $m_s$). This yields the following
simplifications:
\begin{enumerate}
\item[i)] the solution of the classical \eom{} for the eta--field is trivial, $\eta=0$;
\item[ii)] there is no  mixing between the $\eta$ and the $\pi^0$ fields.
\end{enumerate}
Point i) greatly simplifies the transition
from \chpt{3} building blocks of the monomials to those of two flavours, as it
suppresses any effects from the eta, whereas point ii)
 eliminates many possible graphs and hence considerably reduces the 
requested labour. Indeed, in this restricted framework, 
the one--particle reducible graphs (two one--loop diagrams
linked by a single propagator) do not contribute to the
matching: due to strangeness conservation, the linking propagator cannot
be a kaon. Since we can concentrate on local terms only, we can drop the pions
as candidates, too. 
The remaining one--particle reducible diagrams do not contribute to the 
matching at this order, as
can  be verified by working out the algebra of the vertices linking the
single eta propagator with the one--loop part.

Aiming for the $\L_6$-monomials in the generating functional requires rather
many graphs with sunset--like topology. In the two--flavour limit, where one is
interested in the local contributions only, one can simplify the
loop calculations by using a short distance expansion for the massive
propagators. This simplifies drastically the involved loop
integrals; however, the contributions from individual graphs are not chirally
invariant. Collecting terms stemming from different graphs to obtain a
manifestly chirally invariant result is rather cumbersome. Since we are
interested in the local terms only, we use a shortcut which is based 
on gauge invariance\footnote{We
are grateful to H.~Leutwyler for pointing out this possibility to us.}: 
 one may  choose a gauge such that  at some fixed space--time point $x_0$, 
the totally symmetric combination of up
to three derivatives acting on the chiral connection vanish,
\begin{equation}
  \label{eq:choose_gauge}
  \Gamma_\mu(x_0) = 0\co\d_{\{\mu}\Gamma_{\nu\}}(x_0) =
  0\co
  \d_{\{\mu}\d_{\nu}\Gamma_{\rho\}}(x_0) =
  0\co\d_{\{\mu}\d_{\nu}\d_\rho\Gamma_{\sigma\}}(x_0) = 0\fs
\end{equation}
Up to
four ordinary derivatives  are then indistinguishable
from the fully symmetric combinations of covariant derivatives: 
\begin{equation}
\d_{\mu}f(x_0) = \nabla_{\mu}f(x_0)\co
\d_\mu\d_\nu f(x_0) = \tfrac{1}{2}\{\d_\mu,\d_\nu\}f(x_0) = \tfrac12\{ \nabla_\mu,\nabla_\nu \}f(x_0)\co
\text{etc.} 
\end{equation}
This allows us to write even intermediate results in a manifestly chiral
invariant manner. 

To check our calculations, we matched the available $\SU{2}$-- and
$\SU{3}$--results for the vector--vector correlator \cite{Amoros:1999dp} 
(already discussed above) and for the pion form factor, 
worked out in Refs.~
\cite{Bijnens:1998fm} and \cite{Bijnens:2002hp}.
We found that the obtained relations for $\clr{56}$ and
$\clr{51}-\clr{53}$ agree with our findings. Furthermore, we verified the scale
independence of the found relations.

\section{}
\label{sec:result}
\vspace{-2\parskip}
As already stated in Ref.~\cite{Haefeli:2007ty}, the monomial $P_{27}$ can be
discarded from the $p^6$--Lagrangian for \chpt{2}. Therefore, the matching
relations will certainly be a combination of some $\clr{i}$ and $\clr{27}$. 
Due to the restricted framework, only relations for \lecs{} not involving
monomials dependent on the sources $s$ or $p$ are nontrivial. Moreover, in the
restricted framework, there is an additional relation among the remaining
$\SU{2}$--monomials:
\begin{equation}
\begin{split}
&
         \tfrac43P_{1}
       - \tfrac13P_{2}
       + P_{3}
       - \tfrac{10}{3}P_{24}
       + \tfrac43P_{25}
       + 2P_{26}
       - \tfrac83P_{28}
       - \tfrac12P_{29}
\\&
       + \tfrac12P_{30}
       - P_{31}
       + 2P_{32}
       - \tfrac12P_{33}
       + \tfrac43P_{36}
       - \tfrac43P_{37}
       - \tfrac{11}{6}P_{39}
\\&
       + \tfrac56P_{40}
       + \tfrac73P_{41}
       - \tfrac43P_{42}
       - \tfrac32P_{43}
       + \tfrac12P_{44}
       - \tfrac12P_{45}
       - P_{51}
       - P_{53}
=0\fs
\end{split}
\label{eq:new_relation}
\end{equation}
Because the \eom{} is different in the full framework, this relation is no longer
valid there. We used Eq.~(\ref{eq:new_relation})  to exclude
the monomial $P_1$ from our consideration. As a result, we give
the matching for the \osn{27} combinations of $\clr{i}$, as shown in
table~\ref{tab:1}. In the full framework, an additional matching relation
(apart from the ones for the monomials involving the sources $s$ and $p$) for
$\clr{1}$ can be worked out, yielding the only missing piece in the matching for
the \osn{28} \lecs{} worked out here. 

To render the formulae more compact, we found it convenient to
express the bare kaon mass squared $B_0m_s$ through its equivalent $\mk^2$ in
the chiral limit, cf.~(\ref{eq:Mk2}). Then, the final result may be written in the form
\begin{equation}
  \label{eq:p0p1p2}
  x_i = p^{(0)}_i + p^{(1)}_i\ell_K + p^{(2)}_i\ell_K^2+O(m_s)\co
\end{equation}
where $x_i$ denotes one of the \osn{27} linear combinations of the $\clr{i}$
displayed in table~\ref{tab:1}. The explicit expressions for
the polynomials $ p^{(n)}_i$ in the \chpt{3}--\lecs{} are displayed
in tables~\ref{tab:2} and~\ref{tab:3}.
We use the abbreviations
\begin{align}
  Z_s &=\frac{F^2}{16\pi^2 \mk^2}\co\quad
\rho_{1} =\sqrt{2}\,\mathrm{Cl}_2(\arccos(1/3)) 
\sim 1.41602\co\nonumber\\
\mathrm{Cl}_2(\theta)&=-\frac{1}{2}\int_0^\theta
 d\phi \,\, \ln\,(4\sin^2{\frac{\phi}{2}})\fs
\end{align}

\section{}
\label{sec:summary}
\vspace{-2\parskip}
In summary, we have worked out the strange quark mass dependence  of 
two--flavour \lecs{} at order $p^6$. The calculation is performed at two--loop order. 
To simplify the procedure, we have restricted the evaluation to \lecs{} that occur 
in the axial and vector Green functions, in the chiral limit. This 
concerns \osn{28} out of the \osn{56} \lecs{} at this order. The calculation of the 
pertinent relations for the remaining \osn{28} \lecs{}  would require a very 
considerable amount of work.


{\bf Acknowledgments}
\begin{sloppypar}
 We thank H. Leutwyler for very useful discussions.  The Center for
  Research and Education in Fundamental Physics is supported by the
  ``Innovations-- und Kooperationsprojekt C--13'' of the ``Schweizerische
  Universit\"atskonferenz SUK/CRUS''.
 This work was supported by the  Swiss National Science Foundation, 
and by EU MRTN--CT--2006--035482  (FLAVIA{\it net}).
\end{sloppypar}

\vskip2cm


\renewcommand{\arraystretch}{1.5}
\newcommand{\extraline}{$\\ & $}
\setlength{\LTcapwidth}{\textwidth}
\begin{longtable}{rl|rl|rl}
\label{tab:1}

 $i$ &    $x_i$ & $i$ & $x_i$  & $i$ &  $x_i$   \\
\hline\hline
  1 & $ \clr{2}  + \frac{1}{4} \clr{1}$                            &  
 10 & $ \clr{32} - \frac{3}{2} \clr{1} - \clr{27}$                &
 19 & $ \clr{43} + \frac{9}{8} \clr{1} + \frac{1}{4} \clr{27}$      \\
%
  2 & $ \clr{3}  - \frac{3}{4} \clr{1}  $                         &
 11 & $ \clr{33} + \frac{3}{8} \clr{1} + \frac{1}{4} \clr{27}$     & 
 20 & $ \clr{44} - \frac{3}{8} \clr{1}- \frac{1}{4} \clr{27} $      \\
%
  3 & $ \clr{24} + \frac{5}{2} \clr{1}$                           &
 12 & $ \clr{36}- \clr{1}$                                         & 
 21 & $ \clr{45} + \frac{3}{8} \clr{1} + \frac{1}{4} \clr{27} $    \\
%
  4 & $ \clr{25} - \clr{1} $                                      &
 13 & $\clr{37} + \clr{1}$                                         & 
 22 & $ \clr{50} $                                                \\
%
  5 & $ \clr{26} - \frac{3}{2} \clr{1} $                           &
 14 & $ \clr{38}$                                                  & 
 23 & $ \clr{51} + \frac{3}{4} \clr{1}+ \frac{1}{2} \clr{27} $     \\
  6 & $ \clr{28}+ 2 \clr{1} - \clr{27} $                           &
 15 & $ \clr{39} + \frac{11}{8} \clr{1} + \frac{1}{4} \clr{27}$    &
 24 & $ \clr{52}$                                                 \\
%
  7 & $ \clr{29} + \frac{3}{8} \clr{1}  + \frac{1}{4} \clr{27} $    &
 16 & $ \clr{40} - \frac{5}{8} \clr{1} - \frac{1}{4} \clr{27} $     &
 25 & $ \clr{53}+ \frac{3}{4} \clr{1} + \frac{1}{2} \clr{27} $      \\
%
  8 & $ \clr{30} - \frac{3}{8} \clr{1} - \frac{1}{4} \clr{27}$    &
 17 & $ \clr{41} - \frac{7}{4} \clr{1} - \frac{1}{2} \clr{27}$    &
 26 & $ \clr{55}$                                                \\
%
  9 & $\clr{31}+ \frac{3}{4} \clr{1}  + \frac{1}{2} \clr{27}$      &
 18 & $ \clr{42} + \clr{1}$                                        &
 27 & $ \clr{56}$                                                 \\
\caption[]{\rule{0cm}{2ex}The quantities $x_i$ in Eq.~(\ref{eq:p0p1p2})}
\end{longtable}

\newpage

\clearpage
\renewcommand{\multirowsetup}{\flushright}
\begin{longtable}{r|l}
\label{tab:2}
$i$ & $p^{(0)}_i$ \\
\hline\hline
\endfirsthead
\multicolumn{2}{l}{\small\slshape continued from previous page}\\
$i$ & $p^{(0)}_i$ \\
\hline\hline
\endhead
\multicolumn{2}{l}{\small\slshape continued on next page}\\
\endfoot
\endlastfoot
\multirow{2}{1pc}{1} &
$
          - \frac{24271}{589824 N^2}
          - \frac{1}{1920} Z_s
          - \frac{231}{262144 N^2} \ln\tfrac{4}{3}
          + \frac{1}{3 N} \Lr{1}
          + \frac{1}{12 N} \Lr{2}
          + \frac{11}{96 N} \Lr{3}
          - \frac{1}{24 N} \Lr{4}
$\\

&$
          + \frac{1}{4} \Clr{1}
          + \frac{1}{2} \Clr{2}
          + \Clr{3}
          - \frac{2285}{1572864 N^2} \rho_1
$
\\
\hline
\multirow{2}{1pc}{2} &  
$
            \frac{30193}{589824 N^2}
          + \frac{1}{480} Z_s
          + \frac{1099}{786432 N^2} \ln\tfrac{4}{3}
          - \frac{1}{N} \Lr{1}
          - \frac{1}{4 N} \Lr{2}
          - \frac{29}{96 N} \Lr{3}
          + \frac{1}{8 N} \Lr{4}
$\\
&$
          - \frac{3}{4} \Clr{1}
          - \frac{3}{2} \Clr{2}
          + \Clr{4}
          + \frac{5651}{1572864 N^2} \rho_1
$\\
\hline
\multirow{2}{1pc}{3} & 
$
          - \frac{927}{32768 N^2}
          - \frac{1}{576} Z_s
          - \frac{2893}{393216 N^2} \ln\tfrac{4}{3}
          +\frac{1}{N} \Lr{1}
          - \frac{1}{6 N} \Lr{2}
          + \frac{11}{48 N} \Lr{3}
          - \frac{1}{4 N} \Lr{4}
$\\
&$
          + \frac{5}{2} \Clr{1}
          + 5 \Clr{2}
          + \Clr{40}
          + 2 \Clr{41}
          + \Clr{42}
          + \Clr{47}
          - \frac{3223}{262144 N^2} \rho_1 
$\\
\hline
\multirow{2}{1pc}{4} & 
$
          - \frac{18085}{147456 N^2}
          - \frac{3}{640} Z_s
          + \frac{841}{196608 N^2} \ln\tfrac{4}{3}
          + \frac{3}{2 N} \Lr{1}
          - \frac{1}{4 N} \Lr{2}
          + \frac{17}{48 N} \Lr{3}
          - \frac{1}{4 N} \Lr{4}
          - \Clr{1}
$
\\
&$
          - 2 \Clr{2}
          + \Clr{44}
          + 2 \Clr{45}
          + \Clr{47}
          - \frac{1951}{393216 N^2} \rho_1
$\\
\hline
\multirow{2}{1pc}{5} & 
$
            \frac{18091}{98304 N^2}
          + \frac{11}{2880} Z_s
          + \frac{2747}{393216 N^2} \ln\tfrac{4}{3}
          - \frac{2}{N} \Lr{1}
          - \frac{1}{3 N} \Lr{2}
          - \frac{31}{48 N} \Lr{3}
          + \frac{1}{4 N} \Lr{4}
$\\
&$
          - \frac{3}{2} \Clr{1}
          - 3 \Clr{2}
          + \Clr{46}
          - \Clr{47}
          + \frac{8963}{786432 N^2} \rho_1
$\\
\hline
 \multirow{2}{1pc}{6} & 
$
            \frac{6875}{73728 N^2}
          + \frac{7}{480} Z_s
          - \frac{1223}{98304 N^2} \ln\tfrac{4}{3}
           - \frac{10}{3 N} \Lr{1}
          - \frac{1}{3 N} \Lr{2}
          - \frac{3}{4 N} \Lr{3}
          + \frac{2}{3 N} \Lr{4}
$\\
&$
          + \frac{1}{4 N} \Lr{9}
          + 2 \Clr{1}
          + 4 \Clr{2}
          + 2 \Clr{48}
          + 2 \Clr{49}
          - \Clr{50}
          + \Clr{51}
          - \frac{101}{65536 N^2} \rho_1
$\\
\hline
 \multirow{2}{1pc}{7} & 
$
          - \frac{22535}{393216 N^2}
          + \frac{1}{1280} Z_s
          - \frac{2205}{524288 N^2} \ln\tfrac{4}{3}
          + \frac{1}{6 N} \Lr{1}
          - \frac{1}{8 N} \Lr{2}
          + \frac{5}{192 N} \Lr{3}
          + \frac{5}{48 N} \Lr{4}
$\\
&$
          - \frac{1}{8 N} \Lr{10}
          + \frac{3}{8} \Clr{1}
          + \frac{3}{4} \Clr{2}
          + \frac{1}{4} \Clr{50}
          - \frac{1}{4} \Clr{52}
          + \Clr{53}
          + 2 \Clr{54}
          - \frac{5781}{1048576 N^2} \rho_1
$\\
\hline
 \multirow{2}{1pc}{8} & 
$ 
            \frac{260927}{3538944 N^2}
          + \frac{17}{3840} Z_s
          + \frac{4055}{1572864 N^2} \ln\tfrac{4}{3}
          - \frac{1}{2 N} \Lr{1}
          - \frac{1}{24 N} \Lr{2}
          - \frac{29}{192 N} \Lr{3}
          + \frac{1}{16 N} \Lr{4}
$\\
&$
          - \frac{1}{8 N} \Lr{9}
          - \frac{3}{8} \Clr{1}
          - \frac{3}{4} \Clr{2}
          - \frac{1}{4} \Clr{50}
          + \frac{1}{4} \Clr{52}
          + \Clr{55}
          + \frac{35263}{3145728 N^2} \rho_1
$\\
\hline
 \multirow{2}{1pc}{9} & 
$
          - \frac{73685}{589824 N^2}
          + \frac{11}{1920} Z_s
          - \frac{2327}{786432 N^2} \ln\tfrac{4}{3}
          + \frac{1}{N} \Lr{1}
          + \frac{5}{12 N} \Lr{2}
          + \frac{37}{96 N} \Lr{3}
          - \frac{1}{8 N} \Lr{4}
$\\
&$
          - \frac{1}{4 N} \Lr{9}
          + \frac{3}{4} \Clr{1}
          + \frac{3}{2} \Clr{2}
          + \frac{1}{2} \Clr{50}
          - \frac{1}{2} \Clr{52}
          + \Clr{56}
          + \Clr{58}
          + \frac{3841}{1572864 N^2} \rho_1
$\\
\hline
 \multirow{2}{1pc}{10} & 
$
            \frac{6245}{32768 N^2}
          + \frac{1}{192} Z_s
          + \frac{2687}{393216 N^2} \ln\tfrac{4}{3}
          - \frac{2}{N} \Lr{1}
          + \frac{1}{6 N} \Lr{2}
          - \frac{25}{48 N} \Lr{3}
          + \frac{1}{4 N} \Lr{4}
          - \frac{1}{12 N} \Lr{9}
$\\
&$
          - \frac{3}{2} \Clr{1}
          - 3 \Clr{2}
          - \Clr{50}
          + \Clr{52}
          + \Clr{57}
          + \Clr{58}
          + \Clr{60}
          + \frac{3623}{786432 N^2} \rho_1
$\\
\hline
 \multirow{2}{1pc}{11} & 
$
          - \frac{165839}{3538944 N^2}
          - \frac{7}{1280} Z_s
          - \frac{1511}{1572864 N^2} \ln\tfrac{4}{3}
         + \frac{1}{2 N} \Lr{1}
          + \frac{1}{24 N} \Lr{2}
          + \frac{29}{192 N} \Lr{3}
          - \frac{1}{16 N} \Lr{4}
$\\
&$
          + \frac{1}{6 N} \Lr{9}
          + \frac{3}{8} \Clr{1}
          + \frac{3}{4} \Clr{2}
          + \frac{1}{4} \Clr{50}
          - \frac{1}{4} \Clr{52}
          + \Clr{59}
          + \frac{1}{2} \Clr{60}
          - \frac{5455}{3145728 N^2} \rho_1
$\\
\hline
 \multirow{2}{1pc}{12} & 
$
            \frac{30515}{442368 N^2}
          + \frac{1}{192} Z_s
          + \frac{587}{196608 N^2} \ln\tfrac{4}{3}
          - \frac{8}{3 N} \Lr{1}
          - \frac{1}{2 N} \Lr{2}
          - \frac{19}{24 N} \Lr{3}
          + \frac{1}{3 N} \Lr{4}
$\\
&$
          - \frac{1}{24 N} \Lr{9}
          - \Clr{1}
          - 2 \Clr{2}
          + \Clr{66}
          + \frac{1}{2} \Clr{68}
          + \frac{1043}{393216 N^2} \rho_1
$\\
\hline
\multirow{2}{1pc}{13} & 
$
          - \frac{13001}{442368 N^2}
          - \frac{1}{480} Z_s
          - \frac{1009}{196608 N^2} \ln\tfrac{4}{3}
          + \frac{4}{3 N} \Lr{1}
          + \frac{1}{3 N} \Lr{2}
          + \frac{11}{24 N} \Lr{3}
          - \frac{1}{6 N} \Lr{4}
$\\
&$
          + \Clr{1}
          + 2 \Clr{2}
          + \Clr{67}
          + \frac{2359}{393216 N^2} \rho_1
$\\
\caption[]{\rule{0cm}{2ex}The polynomial $p^{(0)}_i$ as defined in Eq.~(\ref{eq:p0p1p2})}\\
\newpage
\multirow{2}{1pc}{14} & 
$
            \frac{3691}{221184 N^2}
          + \frac{1}{480} Z_s
          - \frac{109}{98304 N^2} \ln\tfrac{4}{3}
          - \frac{4}{3 N} \Lr{1}
          - \frac{1}{6 N} \Lr{2}
          - \frac{5}{12 N} \Lr{3}
          + \frac{1}{6 N} \Lr{4}
$\\
&$
          - \frac{1}{12 N} \Lr{9}
          + \frac{1}{2} \Clr{68}
          + \Clr{69}
          + \frac{539}{196608 N^2} \rho_1
$\\
\hline

\multirow{2}{1pc}{15} & 
$
          - \frac{20525}{131072 N^2}
          + \frac{1}{1280} Z_s
          - \frac{17239}{1572864 N^2} \ln\tfrac{4}{3}
          + \frac{1}{2 N} \Lr{1}
          + \frac{1}{24 N} \Lr{2}
          + \frac{29}{192 N} \Lr{3}
          - \frac{1}{16 N} \Lr{4}
$\\
&$
          - \frac{1}{12 N} \Lr{9}
          + \frac{1}{8 N} \Lr{10}
          + \frac{11}{8} \Clr{1}
          + \frac{11}{4} \Clr{2}
          + \frac{1}{4} \Clr{50}
          - \frac{1}{4} \Clr{52}
          + \Clr{70}
          + 2 \Clr{71}
          - \frac{12607}{3145728 N^2} \rho_1
$\\
\hline
\multirow{2}{1pc}{16} & 
$
            \frac{32785}{393216 N^2}
          - \frac{1}{3840} Z_s
          + \frac{4801}{1572864 N^2} \ln\tfrac{4}{3}
          - \frac{5}{6 N} \Lr{1}
          - \frac{1}{8 N} \Lr{2}
          - \frac{17}{64 N} \Lr{3}
          + \frac{5}{48 N} \Lr{4}
 $\\
&$
         + \frac{1}{12 N} \Lr{9}
          - \frac{5}{8} \Clr{1}
          - \frac{5}{4} \Clr{2}
          - \frac{1}{4} \Clr{50}
          + \frac{1}{4} \Clr{52}
          + \Clr{72}
          + \frac{15641}{3145728 N^2} \rho_1
$\\
\hline
 \multirow{2}{1pc}{17} & 
$
            \frac{440347}{1769472 N^2}
          + \frac{1}{384} Z_s
          + \frac{8723}{786432 N^2} \ln\tfrac{4}{3}
          - \frac{7}{3 N} \Lr{1}
          - \frac{5}{12 N} \Lr{2}
          - \frac{73}{96 N} \Lr{3}
          + \frac{7}{24 N} \Lr{4}
$\\
&$
          + \frac{1}{12 N} \Lr{9}
          - \frac{7}{4} \Clr{1}
          - \frac{7}{2} \Clr{2}
          - \frac{1}{2} \Clr{50}
          + \frac{1}{2} \Clr{52}
          + \Clr{73}
          + \Clr{75}
          + \frac{5513}{524288 N^2} \rho_1
$\\
\hline
\multirow{2}{1pc}{18} & 
$
          - \frac{20843}{442368 N^2}
          + \frac{1}{480} Z_s
          - \frac{1763}{196608 N^2} \ln\tfrac{4}{3}
          - \frac{4}{3 N} \Lr{1}
          - \frac{3}{8 N} \Lr{3}
          + \frac{1}{6 N} \Lr{4}
$\\
&$
          - \frac{1}{6 N} \Lr{9}
          + \Clr{1}
          + 2 \Clr{2}
          + \Clr{74}
          + \Clr{75}
          + \Clr{77}
          + \frac{263}{131072 N^2} \rho_1
$\\
\hline
\multirow{2}{1pc}{19} &
$
          - \frac{338965}{3538944 N^2}
          + \frac{1}{1280} Z_s
          - \frac{8989}{1572864 N^2} \ln\tfrac{4}{3}
          + \frac{1}{6 N} \Lr{1}
          + \frac{1}{8 N} \Lr{2}
          + \frac{23}{192 N} \Lr{3}
          - \frac{1}{48 N} \Lr{4}
$\\
&$
          - \frac{1}{12 N} \Lr{9}
          + \frac{9}{8} \Clr{1}
          + \frac{9}{4} \Clr{2}
          + \frac{1}{4} \Clr{50}
          - \frac{1}{4} \Clr{52}
          + \Clr{76}
          + \frac{1}{2} \Clr{77}
          + \frac{4843}{3145728 N^2} \rho_1
$\\
\hline
\multirow{2}{1pc}{20} & 
$
            \frac{7309}{131072 N^2}
          + \frac{7}{1280} Z_s
          + \frac{887}{1572864 N^2} \ln\tfrac{4}{3}
          - \frac{1}{2 N} \Lr{1}
          - \frac{1}{24 N} \Lr{2}
          - \frac{29}{192 N} \Lr{3}
          + \frac{1}{16 N} \Lr{4}
$\\
&$
          - \frac{1}{6 N} \Lr{9}
          - \frac{3}{8} \Clr{1}
          - \frac{3}{4} \Clr{2}
          - \frac{1}{4} \Clr{50}
          + \frac{1}{4} \Clr{52}
          + \Clr{78}
          + \frac{48223}{3145728 N^2} \rho_1 
$\\
\hline
\multirow{2}{1pc}{21} & 
$
          - \frac{58405}{1179648 N^2}
          - \frac{1}{768} Z_s
          + \frac{1657}{1572864 N^2} \ln\tfrac{4}{3}
          + \frac{1}{2 N} \Lr{1}
          + \frac{1}{24 N} \Lr{2}
          + \frac{29}{192 N} \Lr{3}
          - \frac{1}{16 N} \Lr{4}
$\\
&$
          - \frac{1}{12 N} \Lr{9}
          + \frac{3}{8} \Clr{1}
          + \frac{3}{4} \Clr{2}
          + \frac{1}{4} \Clr{50}
          - \frac{1}{4} \Clr{52}
          + \Clr{79}
          - \frac{18415}{3145728 N^2} \rho_1
$\\
\hline
22 & 
$
            \frac{379}{24576 N^2}
          + \frac{1}{480} Z_s
          + \frac{185}{32768 N^2} \ln\tfrac{4}{3}
          - \frac{1}{12 N} \Lr{9}
          + \Clr{87}
          + \frac{81}{65536 N^2} \rho_1 
$\\
\hline
\multirow{2}{1pc}{23} & 
$
          - \frac{70805}{589824 N^2}
          - \frac{1}{384} Z_s
          - \frac{3415}{786432 N^2} \ln\tfrac{4}{3}
           + \frac{1}{N} \Lr{1}
          + \frac{1}{12 N} \Lr{2}
          + \frac{29}{96 N} \Lr{3}
          - \frac{1}{8 N} \Lr{4}
$\\
&$
          + \frac{3}{4} \Clr{1}
          + \frac{3}{2} \Clr{2}
          + \frac{1}{2} \Clr{50}
          - \frac{1}{2} \Clr{52}
          + \Clr{88}
          - \frac{11071}{1572864 N^2} \rho_1
$\\
\hline
24 & 
$
            \frac{1937}{221184 N^2}
          + \frac{1}{96} Z_s
          + \frac{41}{98304 N^2} \ln\tfrac{4}{3}
          + \frac{1}{12 N} \Lr{3}
          - \frac{7}{24 N} \Lr{9}
          + \Clr{89}
          + \frac{875}{65536 N^2} \rho_1
$\\
\hline
\multirow{2}{1pc}{25} & 
$
          - \frac{73357}{589824 N^2}
          - \frac{1}{1920} Z_s
          - \frac{5375}{786432 N^2} \ln\tfrac{4}{3}
          + \frac{1}{N} \Lr{1}
          + \frac{1}{12 N} \Lr{2}
          + \frac{37}{96 N} \Lr{3}
          - \frac{1}{8 N} \Lr{4}
$\\
&$
          - \frac{1}{24 N} \Lr{9}
          + \frac{3}{4} \Clr{1}
          + \frac{3}{2} \Clr{2}
          + \frac{1}{2} \Clr{50}
          - \frac{1}{2} \Clr{52}
          + \Clr{90}
          - \frac{4493}{524288 N^2} \rho_1
$\\
\hline
26 & 
$
            \frac{1}{72 N^2}
          + \frac{1}{90} Z_s
          - \frac{2}{3 N} \Lr{9}
          + \Clr{92}
$\\
\hline
27 & 
$
          - \frac{1}{288 N^2}
          - \frac{1}{240} Z_s
          + \frac{1}{6 N} \Lr{9}
          + \Clr{93}
$\\
\caption[]{\rule{0cm}{2ex}The polynomial $p^{(0)}_i$ as defined in Eq.~(\ref{eq:p0p1p2}) (cont.)}
\end{longtable}


\clearpage

\vspace*{-2cm}
\begin{longtable}{rrr}
\label{tab:3}
\endhead
$i$ &   \hspace*{3.5cm} $p^{(1)}_i$  &  $p^{(2)}_i$ \\
\hline\hline
1 &  
$
          - \frac{913}{27648 N^2} 
          + \frac{1}{4 N}  \Lr{1}
          + \frac{1}{12 N}  \Lr{2}
          + \frac{3}{32 N}  \Lr{3}

$
&$  - \frac{25}{2304 N^2}   
$\\
%
2 &  
$
            \frac{1483}{27648 N^2} 
          - \frac{3}{4 N}  \Lr{1}
          - \frac{1}{4 N}  \Lr{2}
          - \frac{23}{96 N}  \Lr{3}
$
 &$         \frac{5}{256 N^2} 
$\\
%
%
3 & 
$
          - \frac{785}{13824 N^2} 
          + \frac{1}{2 N}  \Lr{1}
          - \frac{1}{6 N}  \Lr{2}
          + \frac{5}{48 N}  \Lr{3}
$
 &$       - \frac{1}{384 N^2} 
$\\
%
4 & 
$
          - \frac{329}{3456 N^2} 
          + \frac{1}{N}  \Lr{1}
          - \frac{1}{4 N}  \Lr{2}
          + \frac{11}{48 N}  \Lr{3}
$
 &$       - \frac{73}{2304 N^2} 
$\\
%
5 & 
$
            \frac{2407}{13824 N^2} 
          - \frac{3}{2 N}  \Lr{1}
          - \frac{1}{3 N}  \Lr{2}
          - \frac{25}{48 N}  \Lr{3}
$
 &$         \frac{11}{288 N^2} 
$\\
\hline
 6 & 
$
            \frac{79}{1152 N^2} 
          - \frac{2}{N}  \Lr{1}
          - \frac{1}{3 N}  \Lr{2}
          - \frac{5}{12 N}  \Lr{3}
          + \frac{1}{4 N}  \Lr{9}
$
 &$         \frac{17}{288 N^2} 
$\\
%
%
 7 & 
$
          - \frac{113}{2048 N^2} 
          + \frac{3}{8 N}  \Lr{1}
          - \frac{1}{8 N}  \Lr{2}
          + \frac{5}{64 N}  \Lr{3}
          - \frac{1}{8 N}  \Lr{10}
$
 &$       - \frac{11}{1536 N^2} 
$\\
%
%
 8 & 
$ 
            \frac{4187}{55296 N^2} 
          - \frac{3}{8 N}  \Lr{1}
          - \frac{1}{24 N}  \Lr{2}
          - \frac{23}{192 N}  \Lr{3}
          - \frac{1}{8 N}  \Lr{9}
$
&$          \frac{61}{4608 N^2} 
$\\
%
%
 9 & 
$
          - \frac{889}{9216 N^2} 
          + \frac{3}{4 N}  \Lr{1}
          + \frac{5}{12 N}  \Lr{2}
          + \frac{31}{96 N}  \Lr{3}
          - \frac{1}{4 N}  \Lr{9}
$
&$        - \frac{19}{768 N^2} 
$\\
%
 10 & 
$
            \frac{737}{4608 N^2} 
          - \frac{3}{2 N}  \Lr{1}
          + \frac{1}{6 N}  \Lr{2}
          - \frac{19}{48 N}  \Lr{3}
          - \frac{1}{12 N}  \Lr{9}
$
&$          \frac{35}{1152 N^2} 
$\\
\hline
 11 & 
$
          - \frac{2395}{55296 N^2} 
          + \frac{3}{8 N}  \Lr{1}
          + \frac{1}{24 N}  \Lr{2}
          + \frac{23}{192 N}  \Lr{3}
          + \frac{1}{6 N}  \Lr{9}
$
&$        - \frac{23}{1536 N^2} 
$\\
%
 12 & 
$
            \frac{605}{6912 N^2} 
          - \frac{2}{N} \Lr{1}
          - \frac{1}{2 N} \Lr{2}
          - \frac{5}{8 N} \Lr{3}
          - \frac{1}{24 N} \Lr{9}
$
&$          \frac{47}{1152 N^2} 
$\\
%
13 & 
$
          - \frac{229}{6912 N^2} 
          + \frac{1}{N} \Lr{1}
          + \frac{1}{3 N} \Lr{2}
          + \frac{3}{8 N} \Lr{3}
$
&$        - \frac{13}{576 N^2} 
$\\
%
14 & 
$
            \frac{29}{864 N^2} 
          - \frac{1}{N} \Lr{1}
          - \frac{1}{6 N} \Lr{2}
          - \frac{1}{3 N} \Lr{3}
          - \frac{1}{12 N} \Lr{9}
$
&$          \frac{5}{288 N^2} 
$\\
%
15 & 
$
          - \frac{707}{6144 N^2} 
          + \frac{3}{8 N}  \Lr{1}
          + \frac{1}{24 N}  \Lr{2}
          + \frac{23}{192 N}  \Lr{3}
          - \frac{1}{12 N}  \Lr{9}
          + \frac{1}{8 N}  \Lr{10}
$
&$        - \frac{73}{4608 N^2} 
$\\
\hline
16 & 
$
            \frac{4109}{55296 N^2} 
          - \frac{5}{8 N}  \Lr{1}
          - \frac{1}{8 N}  \Lr{2}
          - \frac{41}{192 N}  \Lr{3}
          + \frac{1}{12 N}  \Lr{9}
$
&$          \frac{7}{512 N^2} 
$\\
 17 & 
$
            \frac{1997}{9216 N^2} 
          - \frac{7}{4 N}  \Lr{1}
          - \frac{5}{12 N}  \Lr{2}
          - \frac{59}{96 N}  \Lr{3}
          + \frac{1}{12 N}  \Lr{9}
$
&$          \frac{109}{2304 N^2} 
$\\
%
18 & 
$
          - \frac{29}{2304 N^2} 
          - \frac{1}{N} \Lr{1}
          - \frac{7}{24 N}  \Lr{3}
          - \frac{1}{6 N}  \Lr{9}
$
&$          \frac{5}{576 N^2} 
$\\
%
19 & 
$
          - \frac{3745}{55296 N^2} 
          + \frac{1}{8 N}  \Lr{1}
          + \frac{1}{8 N}  \Lr{2}
          + \frac{7}{64 N}  \Lr{3}
          - \frac{1}{12 N}  \Lr{9}
$
&$        - \frac{55}{4608 N^2} 
$\\
%
20 & 
$
            \frac{1273}{18432 N^2} 
          - \frac{3}{8 N}  \Lr{1}
          - \frac{1}{24 N}  \Lr{2}
          - \frac{23}{192 N}  \Lr{3}
          - \frac{1}{6 N}  \Lr{9}
$
&$          \frac{61}{4608 N^2} 
$\\
\hline
21 & 
$
          - \frac{841}{18432 N^2} 
          + \frac{3}{8 N}  \Lr{1}
          + \frac{1}{24 N}  \Lr{2}
          + \frac{23}{192 N}  \Lr{3}
          - \frac{1}{12 N}  \Lr{9}
$
&$        - \frac{37}{4608 N^2} 
$\\
%
22 & 
$
            \frac{19}{1152 N^2} 
          - \frac{1}{12 N}  \Lr{9}
$
&$          \frac{1}{576 N^2} 
$\\
%
23 & 
$
          - \frac{985}{9216 N^2} 
          + \frac{3}{4 N}  \Lr{1}
          + \frac{1}{12 N}  \Lr{2}
          + \frac{23}{96 N}  \Lr{3}
$
&$        - \frac{5}{256 N^2} 
$\\
%
24 & 
$
            \frac{17}{576 N^2} 
          + \frac{1}{12 N}  \Lr{3}
          - \frac{7}{24 N}  \Lr{9}
$
&$          \frac{7}{1152 N^2} 
$\\
%
25 & 
$
          - \frac{3067}{27648 N^2} 
          + \frac{3}{4 N}  \Lr{1}
          + \frac{1}{12 N}  \Lr{2}
          + \frac{31}{96 N}  \Lr{3}
          - \frac{1}{24 N}  \Lr{9}
$
&$          - \frac{43}{2304 N^2} 
$\\
\hline
26 & 
$
            \frac{1}{36 N^2} 
          - \frac{2}{3 N}  \Lr{9}
$
&$          \frac{1}{72 N^2} 
$\\
%
27 & 
$
          - \frac{1}{144 N^2} 
          + \frac{1}{6 N}  \Lr{9}
$
&$          - \frac{1}{288 N^2} 
$\\

\caption[]{\rule{0cm}{2em} The polynomials $p^{(1)}_i$ and $p^{(2)}_i$ as defined in Eq.~(\ref{eq:p0p1p2})}
\end{longtable}


\end{document}